\documentclass[a4paper, 10pt, twocolumn]{article}

\usepackage[ansinew]{inputenc}          
\usepackage[english]{babel}          

\usepackage{graphicx}               
\usepackage{tabularx}               
\usepackage{multirow}               
\usepackage{url}                

\usepackage[small,bf]{caption2}     

\usepackage{psfrag}
\usepackage[printonlyused]{acronym}
\usepackage{xcolor} 

\usepackage[tight]{subfigure}
\usepackage{pgfplots}
\usetikzlibrary{patterns}
\usepgfplotslibrary{units}

\pgfplotsset{compat=1.3}
\pgfplotsset{
  /pgfplots/xbar legend/.style={
  /pgfplots/legend image code/.code={%
  \draw[##1,/tikz/.cd, bar width=3pt,yshift=-0.1em,bar shift=0pt]
  plot coordinates {(0.8em, 0cm) (0.6em, 1.7*\pgfplotbarwidth)};},
  }
}

\usepackage{parskip}

\usepackage{titlesec}
\usepackage{amsmath}                

\titleformat{\section}{\normalfont\large\bfseries}{\thesection}{}{}
\titleformat{\subsection}{\normalfont\large\bfseries}{\thesection}{}{}
\titleformat{\paragraph}{\normalfont\bfseries}{\theparagraph}{}{}
\titlespacing{\section}{0pt}{6pt}{-1pt}
\titlespacing{\subsection}{0pt}{3pt}{-1pt}
\titlespacing{\paragraph}{0pt}{3pt}{-1pt}

\newcolumntype{Y}{>{\centering\arraybackslash}X}    

\acrodef{ASR}{Automatic Speech Recognition}
\acrodef{DOA}{Direction-of-Arrival}
\acrodef{DTFT}{Discrete-Time Fourier Transform}
\acrodef{FIR}{Finite Impulse Response}
\acrodef{FSB}{Filter-and-Sum Beamformer}
\acrodef{fwSegSNR}{frequency-weighted segmental Signal-to-Noise Ratio}
\acrodef{GMM}{Gaussian Mixture Model}
\acrodef{HMM}{Hidden Markov Model}
\acrodef{HRTF}{Head-Related Transfer Function}
\acrodef{LS}{Least-Squares}
\acrodef{MR}{Magnitude Response}
\acrodef{MVDR}{Minimum Variance Distortionless Response}
\acrodef{RIR}{Room Impulse Response}
\acrodef{RLSFI}{Robust Least-Squares Frequency-Invariant}
\acrodef{SIR}{Signal-to-Interference Ratio}
\acrodef{WNG}{White Noise Gain}
\acrodef{WER}{Word Error Rate}

\newcommand{\bb}[1]{\mathbf{#1}}

\newcommand{\mrm}[1]{\mathrm{#1}}

\DeclareMathOperator*{\argmin}{argmin}

\begin{document}

\date{}                                         

\title{\vspace{-8mm}\textbf{\large On the Impact of Localization Errors on HRTF-based Robust Least-Squares Beamforming}}

\author{Hendrik Barfuss and Walter Kellermann\thanks{The research leading to these results has received funding from the European Union's Seventh Framework Programme (FP7/2007-2013) under grant agreement n$^\mathsf{o}$ 609465.}\\
\small Multimedia Communications and Signal Processing,
\small Friedrich-Alexander University Erlangen-N\"urnberg\\
\small \{barfuss,wk\}@lnt.de
}

\maketitle

\thispagestyle{empty}           

\section*{Introduction}
\label{sec:intro}
In a typical human-machine dialogue scenario, the target source and additional interfering sources, located at different positions from the target source, may be active at the same time. Clearly, these interfering sources have to be suppressed in order to establish a successful human-machine interaction. 
A common strategy is to apply spatial filtering techniques which are usually based on the free-field assumption of acoustic wave propagation. However, for scenarios where the microphones are mounted on a scatterer, the free-field assumption is not optimum, since the influence of the scatterer on the sound field is neglected. One example of such a scenario is a microphone array mounted on a robot head used for robot audition, which is also the focus of this article.

In order to design a beamformer which accounts for the influence of the scatterer, i.e., the robot head, on the sound field, the free-field steering vectors have to be replaced by \acp{HRTF}\footnote{Note that in the context of this work, \acp{HRTF} only model the direct propagation path between a source and a microphone mounted on a robot head, but no reverberation components.}, see, e.g., \cite{Maazaoui_Journal_EURASIP2012}.
 
In \cite{Barfuss_WASPAA:2015}, we proposed an \ac{HRTF}-based \ac{RLSFI} beamformer design and verified experimentally that employing \acp{HRTF} instead of free-field steering vectors leads to a significantly improved beamforming performance and correspondingly better \ac{ASR}, in a robot audition scenario.
%
Since the proposed beamformer design depends on a set of \acp{HRTF}, the question arises how the beamformer performs if these \acp{HRTF} do not correspond to the true position of the target source, e.g., due to localization errors. Therefore, in this contribution, we investigate the impact of localization errors on the performance of the \ac{HRTF}-based \ac{RLSFI} beamformer.

The remainder of this article is organized as follows: In the next section, the \ac{HRTF}-based beamformer design from \cite{Barfuss_WASPAA:2015} is briefly reviewed. After this, the results of our investigation of the \ac{HRTF} robustness are presented, followed by a conclusion and an outlook to future work in the last section.

\section*{HRTF-based robust beamforming}
\label{sec:RLSFIbeamforming}

\begin{figure}[t]
  \centering
  \scriptsize
  \psfrag{x0}[cr][cr]{$x_{0}[k]$}
  \psfrag{x1}[cr][cr]{$x_{1}[k]$}
  \psfrag{xN}[cr][cr]{$x_{N-1}[k]$}
  \psfrag{y}[cl][cl]{$y[k]$}   
  \psfrag{w0}[c][c]{$\bb{w}_{0}$}
  \psfrag{w1}[c][c]{$\bb{w}_{1}$}
  \psfrag{wN}[c][c]{$\bb{w}_{N-1}$} 
  \psfrag{p1}[cl][cl]{$\bb{p}_{0}$}
  \psfrag{p2}[cl][cl]{$\bb{p}_{1}$}
  \psfrag{pP}[cl][cl]{$\bb{p}_{N-1}$}    
  \psfrag{+}[c][c]{$+$}    
  \includegraphics[scale = .95]{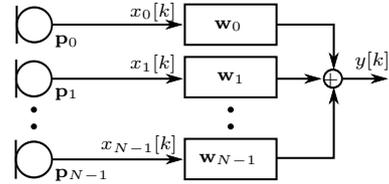}
  \caption{Block diagram of a filter-and-sum beamformer consisting of $N$ microphones and \ac{FIR} filters \cite{Barfuss_WASPAA:2015,lnt2009-22}.}
  \label{fig:FSB}  
\end{figure}

Fig.~\ref{fig:FSB} illustrates the block diagram of a \ac{FSB}, consisting of $N$ microphones at positions $\bb{p}_{n}$, where $\bb{p}_{n}$ represents the position of the $n$-th microphone in Cartesian coordinates. In this article, vectors and matrices are denoted by lower- and upper-case boldface letters, respectively. The output signal $y[k]$ at time instant $k$ is obtained by convolving the microphone signals $x_n[k]$ with \ac{FIR} filters $\bb{w}_{n} = [w_{n,0}, \ldots, w_{n,L-1}]^{T}$ of length $L$ and a subsequent summation over all $N$ channels.
The beamformer response of an \ac{FSB} is given as \cite{VanTrees:2004,lnt2009-22}:
\begin{equation}
  B(\omega, \phi, \theta) = \sum\limits_{n=0}^{N-1} W_{n}(\omega) g_{n}(\omega, \phi, \theta),
  \label{eq:BFResponse}
\end{equation}
where $W_{n}(\omega) = \sum_{l=0}^{L-1} w_{n,l} e^{-j \omega l}$ is the \ac{DTFT} representation of $\mathbf{w}_{n}$. Moreover, $g_{n}(\omega, \phi, \theta)$ is the response of the $n$-th microphone to a plane wave with frequency $\omega$ traveling in the direction $(\phi, \theta)$, where $\phi$ and $\theta$ denote azimuth and elevation angle, respectively, and are defined as in \cite{VanTrees:2004}.

In \cite{lnt2009-22}, the design of an \ac{RLSFI} \ac{FSB} was proposed, where a desired beamformer response $\hat{B}(\omega, \phi, \theta)$ is approximated in the \ac{LS} sense at each frequency $\omega$ subject to a distortionless response constraint in the desired look direction and a constraint on the \ac{WNG}. The \ac{LS} approximation is performed for a discrete set of $P$ frequencies $\omega_{p}$ and $M$ look directions $(\phi_{m},\theta_{m})$, and can be formulated in matrix notation as\footnote{A MATLAB design tool with a graphical user interface for the free-field-based design can be downloaded from \textcolor{blue}{\url{http://goo.gl/obnZwY}}.} \cite{lnt2009-22}
\begin{equation}
  \argmin\limits_{\bb{w}_\mathrm{f}(\omega_{p})} \Vert \bb{G}(\omega_{p}) \bb{w}_\mathrm{f}(\omega_{p}) - \hat{\bb{b}} \Vert_{2}^{2}
  \label{eq:OP_1}
\end{equation}
subject to constraints on the WNG and the response in desired look direction, respectively:
\begin{equation}
  \frac{ |\bb{w}^{T}_\mathrm{f}(\omega_{p}) \bb{d}(\omega_{p})|^{2}}{  \bb{w}^{H}_\mathrm{f}(\omega_{p}) \bb{w}_\mathrm{f}(\omega_{p}) } \ge \gamma > 0, \quad  \bb{w}^{T}_\mathrm{f}(\omega_{p}) \bb{d}(\omega_{p}) = 1,
  \label{eq:OP_2}
\end{equation}
where $\bb{w}_\mathrm{f}(\omega_{p}) \!\! = \!\! [W_{0}(\omega_{p}), \ldots, W_{N-1}(\omega_{p})]^{T}$, $\displaystyle [\bb{G}(\omega_{p})]_{mn} \!\! = \!\! g_{n}(\omega_{p},\phi_{m},\theta_{m})$, vector $\displaystyle \hat{\bb{b}} = [\hat{B}(\phi_{0},\theta_{0}), \ldots, \hat{B}(\phi_{M-1},\theta_{M-1})]^{T}$ contains the desired responses for all $M$ discrete look directions, and $\displaystyle \bb{d}(\omega_{p}) = [g_{0}(\omega_{p},\phi_\mrm{d},\theta_\mrm{d}), \ldots, g_{N-1}(\omega_{p},\phi_\mrm{d},\theta_\mrm{d})]^{T}$ is the steering vector corresponding to the desired look direction $(\phi_\mrm{d}, \theta_\mrm{d})$.
Operators $\Vert \cdot \Vert_{2}$, $(\cdot)^{T}$, and $(\cdot)^{H}$ denote the Euclidean norm, and the transpose and conjugate transpose of vectors or matrices, respectively. Note that the same desired response is chosen for all frequencies, as can be seen from the frequency-independent entries of $\hat{\bb{b}}$.
Equations (\ref{eq:OP_1}) and (\ref{eq:OP_2}) can be interpreted as follows: The \ac{LS} approximation of the desired beamformer response is given by (\ref{eq:OP_1}). The first part of (\ref{eq:OP_2}) represents the \ac{WNG} constraint, with the lower bound $\gamma$ on the \ac{WNG}, which has to be defined by the user. The second part of (\ref{eq:OP_2}) describes the distortionless response constraint which ensures that the target signal, coming from the desired look direction, passes the beamformer undistorted.
The time-domain \ac{FIR} filters $\bb{w}_{n}$ are obtained by solving (\ref{eq:OP_1}), (\ref{eq:OP_2}) for each frequency $\omega_{p}$ separately, followed by an \ac{FIR} approximation of the optimum filter coefficients.

Assuming the microphones are located in the free field, the sensor response is given as 
\begin{equation}
  g_{n,\mrm{FF}}(\omega_{p}, \phi_{m}, \theta_{m}) = e^{-j\mathbf{k}^{T}(\omega_{p}, \phi_{m}, \theta_{m}) \mathbf{p}_{n}},
  \label{eq:g_FF}
\end{equation}
where $\mathbf{k}(\omega_{p}, \phi_{m}, \theta_{m})$ denotes the wave vector which depends on the current frequency and look direction, and the speed of sound \cite{VanTrees:2004}. Thus, matrix $\bb{G}(\omega_{p})$ in (\ref{eq:OP_1}) contains the well-known free-field-based steering vectors with respect to the $M$ look directions and the $N$ microphones, and vector $\bb{d}(\omega_{p})$ in (\ref{eq:OP_2}) is the free-field-based steering vector corresponding to the desired look direction.

The \ac{HRTF}-based \ac{RLSFI} beamformer design, as proposed in \cite{Barfuss_WASPAA:2015}, is obtained by including measured or simulated \acp{HRTF} in (\ref{eq:OP_1}) and (\ref{eq:OP_2}) instead of free-field-based steering vectors. In this case, the sensor response is given as
\begin{equation}
  g_{n,\mrm{HRTF}}(\omega_{p}, \phi_{m}, \theta_{m}) = h_{mn}(\omega_{p}),
  \label{eq:g_HRTF}
\end{equation}
where $h_{mn}(\omega_{p})$ is the \ac{HRTF} modeling the propagation between the $m$-th source position and $n$-th sensor at frequency $\omega_{p}$. Consequently, $\bb{G}(\omega_{p})$ now consists of all \acp{HRTF} between the $M$ look directions and the $N$ microphones, and $\bb{d}(\omega_{p})$ contains the \acp{HRTF} corresponding to the desired look direction. Note that in contrast to the free-field-based design (\ref{eq:g_FF}), the \acp{HRTF}-based design implicitly depends on the robot-source distance for which the \acp{HRTF} have been measured (see, e.g., \cite{Blauert:1997}).

\begin{figure}[t]
\subfigure{
  \hspace{8mm}
  \begin{tikzpicture}[scale=1,trim axis left]
  \node at (-0.975,1.575) {\scriptsize (a)};
  \node at (6.75,1.875) {\scriptsize $\mathrm{BP}_\mathrm{dB}$ [dB]};    
    \begin{axis}[
      label style = {font=\scriptsize},
      tick label style = {font=\tiny},   
      ylabel style={yshift=-1mm}, 
      width=8.91cm,height=3.25cm,grid=major,grid style = {dotted,black},  		
      axis on top, 	
      enlargelimits=false,
      xmin=300, xmax=5000, ymin=0, ymax=180,
      xtick={300,1000,2000,3000,4000,5000},
      xticklabels={\empty},
      ytick={0,45,90,135,180},
      ylabel={$\phi\, [^\circ]\,\rightarrow$},
      colorbar horizontal, colormap/jet, 
      colorbar style={
	at={(0,1.15)}, anchor=north west, font=\tiny, width=6cm, height=0.15cm, xticklabel pos=upper
      },
      point meta min=-40, point meta max=0]
      \addplot graphics [xmin=270, xmax=5035, ymin=0, ymax=185] {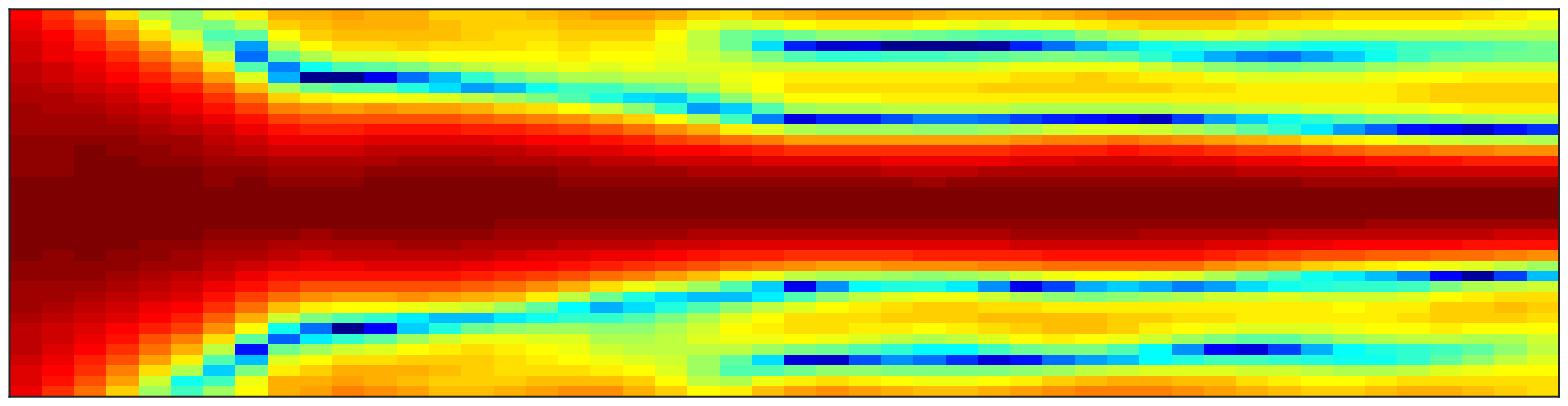};
    \end{axis}
  \end{tikzpicture}
  }
  \\[-4mm]       
  \subfigure{	
  \hspace{8mm}
  \begin{tikzpicture}[scale=1,trim axis left]
  \node at (-0.975,1.575) {\scriptsize (b)};
    \begin{axis}[
      label style = {font=\scriptsize},
      tick label style = {font=\tiny},   
      ylabel style={yshift=-1mm},  	 
      width=8.91cm,height=3.25cm,grid=major,grid style = {dotted,black},  		
      axis on top, 	
      enlargelimits=false,
      xmin=300, xmax=5000, ymin=0, ymax=180,
      xtick={300,1000,2000,3000,4000,5000},
      xticklabels={\empty},
      ytick={0,45,90,135,180},
      ylabel={$\phi\, [^\circ]\,\rightarrow$}]
      \addplot graphics [xmin=270, xmax=5035, ymin=0, ymax=185] {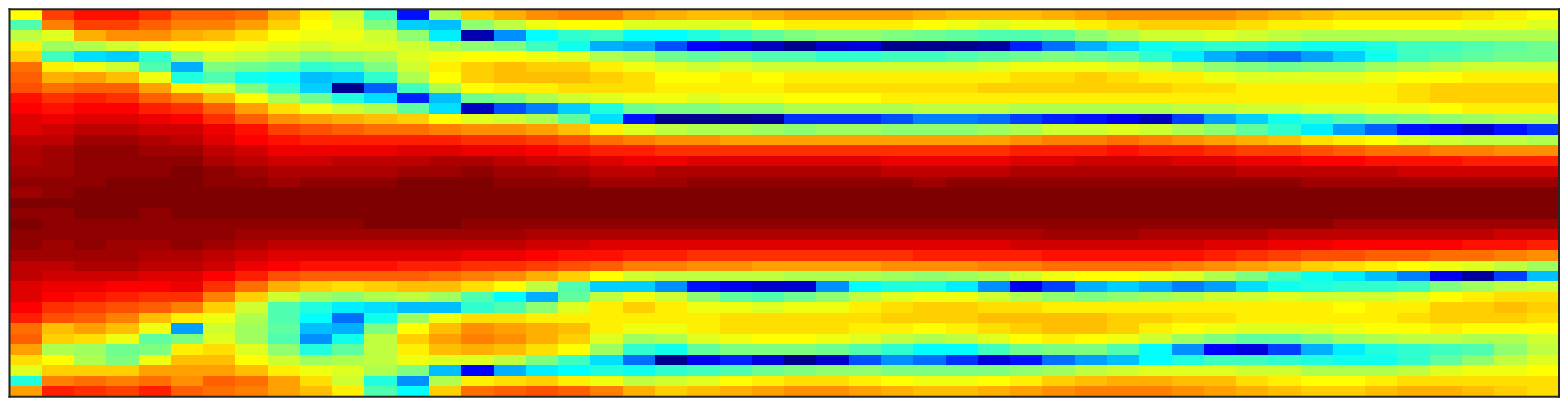};
    \end{axis}
  \end{tikzpicture} 
  }\\[-3.5mm]
  \subfigure{  
  \hspace{8mm}
    \begin{tikzpicture}[scale=1,trim axis left]
    \node at (-0.975,1.575) {\scriptsize (c)};
    \begin{axis}[
      label style = {font=\scriptsize},
      tick label style = {font=\tiny},
      ylabel style={yshift=-2mm},  	
      legend style={font=\tiny, yshift=0.25mm, at={(.6875,0.97)}},
      legend columns = -1,    
      width=8.91cm,height=3.25cm,grid=major,grid style = {dotted,black},
      xtick={300,1000,2000,3000,4000,5000},
      xticklabels={$300$,$1000$,$2000$,$3000$,$4000$,$5000$},
      xlabel={$f\, [\mathrm{Hz}] \, \rightarrow$},	
      ytick={-20, -15, -10, -5, 0},  
      ylabel={$\text{WNG}\, [\text{dB}] \,\rightarrow$},
      ymin=-22.5, ymax=2.5, xmin=300,xmax=5000]   
      \addplot[thick,blue,solid] table [x index=0, y index=2]{WNG_hrtf_WASPAASetup_HRTFnormLookDir_nmic_5_az_90_WNGlim_-20_-10.dat}; \addlegendentry{$\gamma_\mathrm{dB}=-10\, \text{dB}\,\,\,\,$};
      \addplot[thick,red,dashed] table [x index=0, y index=1]{WNG_hrtf_WASPAASetup_HRTFnormLookDir_nmic_5_az_90_WNGlim_-20_-10.dat}; \addlegendentry{$\gamma_\mathrm{dB}=-20\, \text{dB}$};	
    \end{axis}       
  \end{tikzpicture}    
  }
  \vspace{-7mm}
  \caption{Design example of an \ac{HRTF}-based \ac{RLSFI} beamformer, designed for the $5$-microphone robot head array in Fig.~\ref{fig:setup_headArray}. Beampatterns for \ac{WNG} constraints are illustrated in (a) $\gamma_\mathrm{dB}=-10\, \text{dB}$ and (b) $\gamma_\mathrm{dB}=-20\, \text{dB}$. Subfigure (c) shows the resulting \ac{WNG}.}
  \label{fig:designexample_hrtf}
  \vspace{-4mm}
\end{figure}

In Fig.~\ref{fig:designexample_hrtf}, an example of the \ac{HRTF}-based \ac{RLSFI} beamformer according to (\ref{eq:OP_1}), (\ref{eq:OP_2}), and (\ref{eq:g_HRTF}) is illustrated for a frequency range of $300\,\mathrm{Hz} \leq f \leq 5000\,\mathrm{Hz}$. The design was carried out for the $5$-microphone robot head array illustrated in Fig~\ref{fig:setup_headArray}. Beampatterns for two different \ac{WNG} constraint values $\gamma_\mathrm{dB} = 10\log_{10}(\gamma) \in \{-10, -20\}$dB are shown to demonstrate the impact of the \ac{WNG} constraint on the beamformer. It is important to note that the beampatterns were computed by evaluating (\ref{eq:BFResponse}) with (\ref{eq:g_HRTF}). Thus, they effectively show the transfer function between source position and beamformer output with \acp{HRTF} modeling the acoustic system.
We used a filter length of $L=1024$ for the \ac{FIR} approximation, and the \acp{HRTF} which were incorporated in the beamformer design were measured for a robot-source distance of $1.1$m. The main beam was steered towards broadside. 
Figs.~\ref{fig:designexample_hrtf}(a) and \ref{fig:designexample_hrtf}(b) illustrate the resulting beampatterns $\mathrm{BP}_\mathrm{dB}(\ldots) = 10\log_{10}(|B(\ldots)|^{2})$ in dB and Fig.~\ref{fig:designexample_hrtf}(c) shows the corresponding \ac{WNG} in dB over frequency.
It can be seen that both beamformers exhibit a good spatial selectivity above $1000$Hz, and that a higher \ac{WNG} constraint $\gamma_\mathrm{dB}$ leads to a broader beam at lower frequencies. Thus, the user can control the trade-off between robustness and spatial selectivity directly.
Fig~\ref{fig:designexample_hrtf}(c) confirms that both designs fulfill the required \ac{WNG} with occasional slight deviations, which are due to the \ac{FIR} approximation of the optimum filter coefficients. 
Note that a comparison of the beampatterns of the \ac{HRTF}- and free-field-based beamformer with \acp{HRTF} modeling the acoustic system can be found in \cite{Barfuss_WASPAA:2015}, illustrating the effect of the robot head as scatterer on the sound field. 

\section*{Experimental results}
\label{sec:Experiments}
In the following, we analyze the relative robustness of the \ac{HRTF}-based beamformer design by comparing the impact of localization errors on the performance of the \ac{HRTF}- and free-field-based \ac{RLSFI} beamformer. More specifically, we investigate the impact of localization errors with respect to \ac{DOA} and robot-source distance. At first, the experimental setup and performance measures are introduced, followed by a presentation of the experimental results.

\subsection*{Setup and parameters}
We use \acfp{WER} of an automatic speech recognizer to evaluate the overall quality of the enhanced signals at the beamformer outputs, since a high speech recognition accuracy is the main goal in robot audition. As \ac{ASR} engine, we employed PocketSphinx \cite{Huggins:2006} with a \ac{HMM}-\ac{GMM}-based acoustic model trained on clean speech from the GRID corpus \cite{Cooke:2006}, using MFCC+$\Delta$+$\Delta \Delta$ features and cepstral mean normalization. For the computation of the \ac{WER} scores, only the letter and the number in the utterance were evaluated, as in the CHiME challenge \cite{ChristensenBMG10}. Our test signal contained $200$ utterances.
In addition, the \ac{fwSegSNR} as defined in \cite{Hu:2008} was evaluated, where the target signal at the center microphone and at the beamformer output was used as reference signal for calculating the \ac{fwSegSNR} at the input and output of the beamformer, respectively.

\begin{figure}[b]
  \subfigure[Source positions.]{
    \psfrag{inta}[cl][cl]{\parbox[t]{2cm}{\scriptsize \color{red} 1) $\phi_\mrm{int}=70^\circ$}}
    \psfrag{intb}[cl][cl]{\parbox[t]{2cm}{\scriptsize \color{red} 2) $\phi_\mrm{int}=170^\circ$}}
    \psfrag{t}[c][c]{\scriptsize \color{green} target}
    \psfrag{20}[c][c]{\scriptsize $20^\circ$}
    \psfrag{d}[cr][cr]{\scriptsize $1.1\, \text{m}$}
    \psfrag{x1}[cl][cl]{\scriptsize $x_{N-1}$}
    \psfrag{x0}[cl][cl]{\scriptsize $x_0$}
    \includegraphics[width = 3cm]{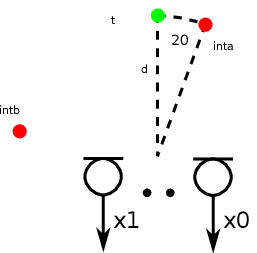}
    \label{fig:evaluation_scenarios}
  }
  \hfill   
  \subfigure[Microphone positions.]{    
    \includegraphics[width = 4cm]{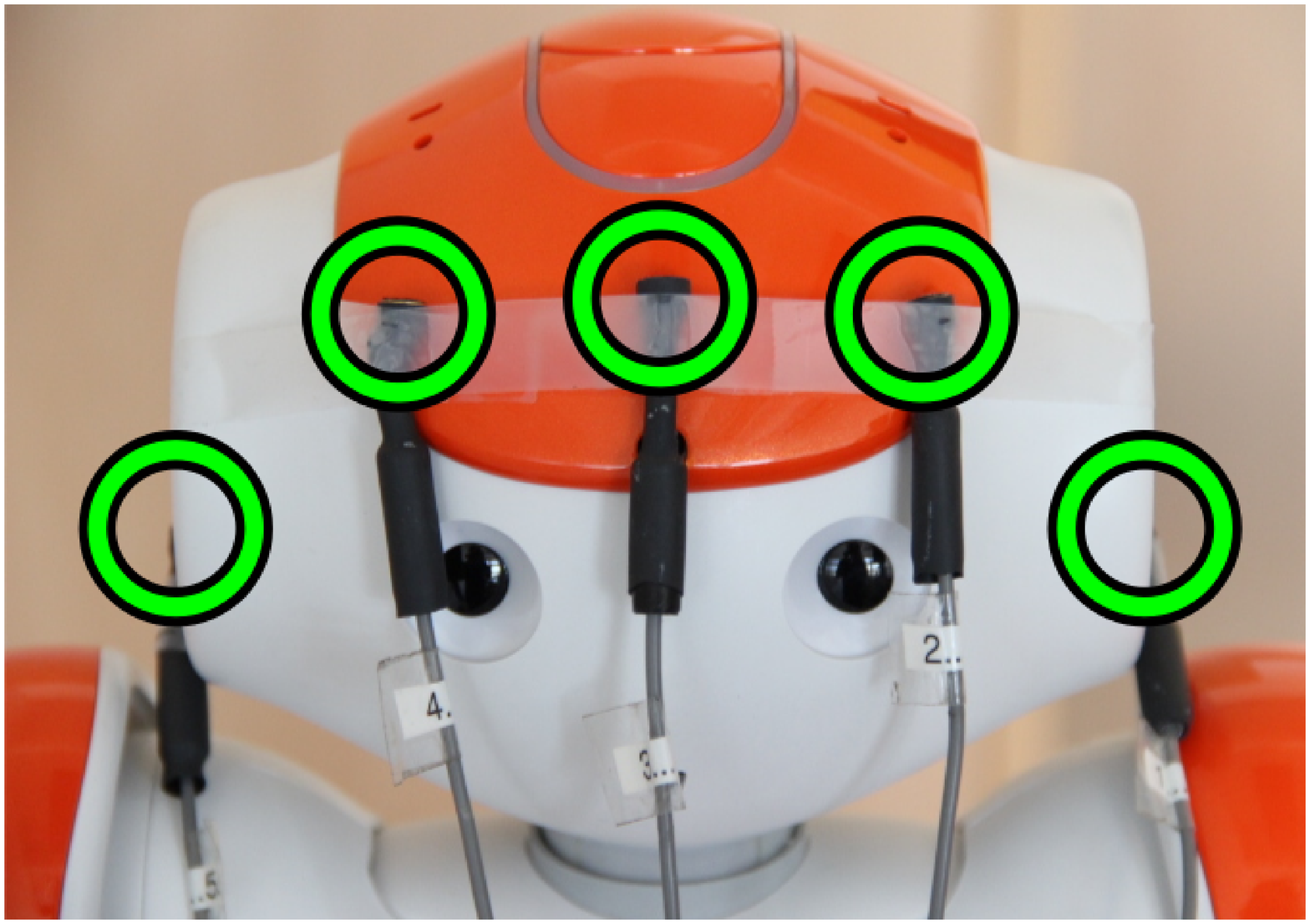}
    \label{fig:setup_headArray}
  }
  \vspace{-3mm}
  \caption{Illustration of the source positions of the two-speaker scenario and the employed microphone positions at the robot head.}
  \vspace{-3mm}
\end{figure}

We created the microphone signals by convolving clean-speech source signals with \acp{RIR}, measured in a lab room with a reverberation time of $T_{60} = 190\, \mathrm{ms}$ and a critical distance 
of approximately $1.2 \, \text{m}$, using maximum-length sequences. The sampling rate of the speech signals and measured \acp{RIR} and \acp{HRTF} was $16$kHz. The microphone positions at the robot head for which the \acp{RIR} were measured are illustrated in Fig.~\ref{fig:setup_headArray}. The relative height of the source with respect to the robot head was $0.73\, \text{m}$, corresponding to an elevation angle $\theta = 56.4^\circ$. This setup was chosen to simulate a taller human interacting with the NAO robot of height $0.57 \, \text{m}$. The measurements were carried out for the robot looking towards broadside.

The set of \acp{HRTF} which is required for the \ac{HRTF}-based beamformer design was measured for the same microphone configuration and robot-source distance as for the \ac{RIR} measurements described above. 

The \ac{HRTF}- and free-field-based beamformers were designed for a filter length of $L=1024$ taps and a \ac{WNG} constraint with a lower bound of $\gamma_\mathrm{dB}-10$ dB.

\subsection*{Impact of localization errors with respect to direction of arrival}
At first, the impact of \ac{DOA} estimation errors on the beamforming performance is investigated. To this end, two two-speaker scenarios were evaluated, where the target source was always located at $\phi_\mrm{d} = 90^\circ$ and the interfering source was located at 1) $\phi_\mrm{int}=70^\circ$ or 2) $\phi_\mrm{int}=170^\circ$, at a robot-source distance of $d=1.1$m. The beamformer was steered towards $\phi_\mrm{BF} \in \{100^\circ, 95^\circ, 90^\circ, 85^\circ, 80^\circ\}$, simulating localization errors of $\pm 5^\circ$ and $\pm 10^\circ$. The scenario was chosen to analyze the impact of localization errors on the beamformer performance in situations where an interfering source is 1) very close to or 2) relatively far away from the target source which is located directly in front of the robot. The evaluated two-speaker scenarios 1) and 2) are illustrated in Fig.~\ref{fig:evaluation_scenarios}, where target source and interfering source positions are illustrated by green and red filled circles, respectively. 

%
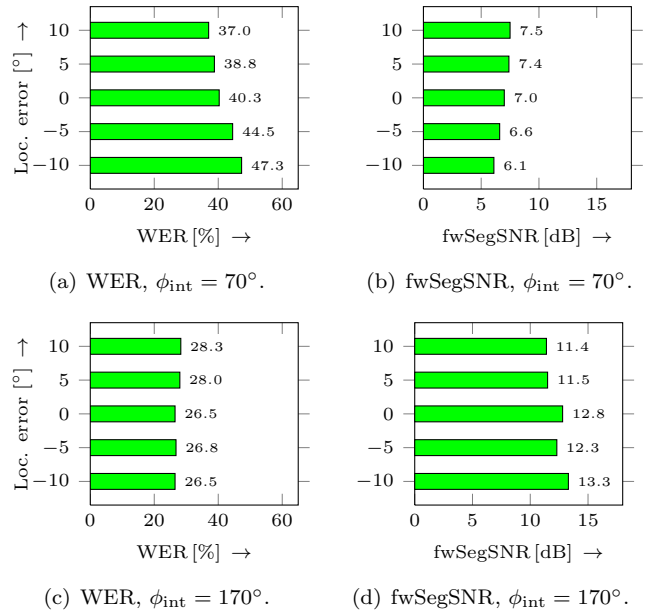
\begin{figure}[t]
  \subfigure[$\text{WER}$, $\phi_\mrm{int}=70^\circ$.]{
    \begin{tikzpicture}
      \begin{axis}[width=0.24\textwidth, height=4cm, 
	label style = {font=\scriptsize},	
	tick label style = {font=\scriptsize},   
	ylabel style={yshift=-2.5mm}, 
	xbar=0pt,
	xmin=0, xmax = 65, ymin=-13.5, ymax=13.5,
	ytick=data,
	xtick={0, 20, 40, 60},
	ylabel = {Loc. error $[^\circ] \, \rightarrow$},
	xlabel = {$\text{WER}\, [\%] \, \rightarrow$},
	bar width = 6pt, 
	every node near coord/.append style={font=\tiny,/pgf/number format/.cd,
            fixed,
            fixed zerofill,
            precision=1,
	    /tikz/.cd},
	nodes near coords, nodes near coords align={horizontal},
	]
	\addplot+[xbar, no markers, black, fill=green] plot coordinates {(47.3,-10) (44.5,-5) (40.3,0) (38.8,5) (37.0,10)};
      \end{axis}
    \end{tikzpicture}
    \label{fig:results_DOAError_HRTF_a_WER}
  }
  \hfill
  \subfigure[$\text{fwSegSNR}$, $\phi_\mrm{int}=70^\circ$.]{
    \begin{tikzpicture}
      \begin{axis}[width=0.24\textwidth, height=4cm, 
	label style = {font=\scriptsize},	
	tick label style = {font=\scriptsize},   
	ylabel style={yshift=-2.5mm}, 
	xbar=0pt,
	xmin=0, xmax = 18, ymin=-13.5, ymax=13.5,
	ytick=data,
	xtick={0, 5, 10, 15},
	xlabel = {$\text{fwSegSNR}\, [\text{dB}] \, \rightarrow$},
	bar width = 6pt, 
	every node near coord/.append style={font=\tiny,/pgf/number format/.cd,
            fixed,
            fixed zerofill,
            precision=1,
	    /tikz/.cd},
	nodes near coords, nodes near coords align={horizontal},
	]
	\addplot+[xbar, no markers, black, fill=green] plot coordinates {(6.1,-10) (6.6,-5) (7.0,0) (7.4,5) (7.5,10)};
      \end{axis}
    \end{tikzpicture}
    \label{fig:results_DOAError_HRTF_a_FWSEGSNR}
  }

  \subfigure[$\text{WER}$, $\phi_\mrm{int}=170^\circ$.]{
    \begin{tikzpicture}
      \begin{axis}[width=0.24\textwidth, height=4cm, 
	label style = {font=\scriptsize},	
	tick label style = {font=\scriptsize},   
	ylabel style={yshift=-2.5mm}, 
	xbar=0pt,
	xmin=0, xmax = 65, ymin=-13.5, ymax=13.5,
	ytick=data,
	xtick={0, 20, 40, 60},
	ylabel = {Loc. error $[^\circ] \, \rightarrow$},
	xlabel = {$\text{WER}\, [\%] \, \rightarrow$},
	bar width = 6pt, 
	every node near coord/.append style={font=\tiny,/pgf/number format/.cd,
            fixed,
            fixed zerofill,
            precision=1,
	    /tikz/.cd},
	nodes near coords, nodes near coords align={horizontal},
	]
	\addplot+[xbar, no markers, black, fill=green] plot coordinates {(26.5,-10) (26.8,-5) (26.5,0) (28.0,5) (28.3,10)};
      \end{axis}
    \end{tikzpicture}
    \label{fig:results_DOAError_HRTF_b_WER}
  }
  \hfill
  \subfigure[$\text{fwSegSNR}$, $\phi_\mrm{int}=170^\circ$.]{
    \begin{tikzpicture}
      \begin{axis}[width=0.24\textwidth, height=4cm, 
	label style = {font=\scriptsize},	
	tick label style = {font=\scriptsize},   
	ylabel style={yshift=-2.5mm}, 
	xbar=0pt,
	xmin=0, xmax = 18, ymin=-13.5, ymax=13.5,
	ytick=data,
	xtick={0, 5, 10, 15},
	xlabel = {$\text{fwSegSNR}\, [\text{dB}] \, \rightarrow$},
	bar width = 6pt, 
	every node near coord/.append style={font=\tiny,/pgf/number format/.cd,
            fixed,
            fixed zerofill,
            precision=1,
	    /tikz/.cd},
	nodes near coords, nodes near coords align={horizontal},
	]
	\addplot+[xbar, no markers, black, fill=green] plot coordinates {(13.3,-10) (12.3,-5) (12.8,0) (11.5,5) (11.4,10)};
      \end{axis}
    \end{tikzpicture}
    \label{fig:results_DOAError_HRTF_b_FWSEGSNR}
  }
  \vspace{-3mm}
  \caption{Illustration of \acp{WER} in $\%$ and \ac{fwSegSNR} levels in $\text{dB}$, obtained at the output of the \ac{HRTF}-based beamformer for Scenario 1) $\phi_\mrm{d}=90^\circ$, $\phi_\mrm{int}=70^\circ$ and 2) $\phi_\mrm{d}=90^\circ$, $\phi_\mrm{int}=170^\circ$ for \ac{DOA} estimation errors of $\pm 5^\circ$ and $\pm 10^\circ$. Measures at input: Scenario 1) $\mathrm{WER}_\mrm{in}=49.0\%$ and $\mathrm{fwSegSNR}_\mrm{in}=5.2$dB and Scenario 2) $\mathrm{WER}_\mrm{in}=44.3\%$ and $\mathrm{fwSegSNR}_\mrm{in}=5.8$dB.}
  \label{fig:results_DOAError_HRTF_a_b}
  \vspace{-4mm}
\end{figure}
In Fig.~\ref{fig:results_DOAError_HRTF_a_b}, the results for the two scenarios are summarized. The subfigures on the left- and right-hand side show the \acp{WER} in $\%$ and \ac{fwSegSNR} levels in dB obtained at the \ac{HRTF}-based beamformer output, respectively. Each horizontal bar represents the results for one specific localization error of $\pm 5^\circ$ or $\pm 10^\circ$.
From Figs.~\ref{fig:results_DOAError_HRTF_a_WER} and \ref{fig:results_DOAError_HRTF_a_FWSEGSNR} it can be seen that when the interferer is very close to the target source, localization errors have a strong impact on the beamforming performance. When the beamformer is accidentally steered closer towards the interfering source (localization errors of $-5^\circ$ and $-10^\circ$), the beamforming performance decreases. This is because the beamformer's main beam is steered towards the interfering source, leading to a lower attenuation of the latter.
If the localization error leads to the beamformer being steered away from the interfering source (localization errors of $5^\circ$ and $10^\circ$), an increasing beamforming performance can be observed. This can be explained by the fact that in this particular scenario, a spatial null of the beampattern is getting closer to the interferer's direction the larger the localization error is.
If the interferer is far away from the target source, as in Scenario 2), localization errors do not have a strong impact on the beamforming performance, which can be seen in Figs.~\ref{fig:results_DOAError_HRTF_b_WER} and \ref{fig:results_DOAError_HRTF_b_FWSEGSNR}, respectively. 

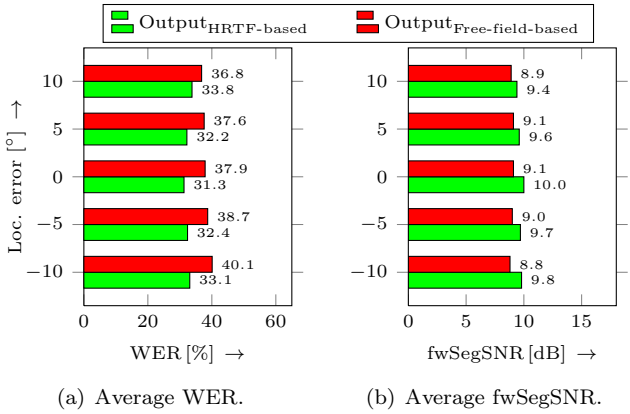
\begin{figure}[t]
  {\hspace{15mm}\ref{legendResults}}\vspace{-1.5mm}  
  \subfigure[Average $\text{WER}$.]{
    \begin{tikzpicture}
      \begin{axis}[width=0.24\textwidth, height=5cm, 
	label style = {font=\scriptsize},	
	tick label style = {font=\scriptsize},   
	ylabel style={yshift=-2.5mm}, 
	xbar=0pt,
	xmin=0, xmax = 65, ymin=-13.5, ymax=13.5,
	ytick=data,
	xtick={0, 20, 40, 60},
	ylabel = {Loc. error $[^\circ] \, \rightarrow$},
	xlabel = {$\text{WER}\, [\%] \, \rightarrow$},
	bar width = 6pt, 
	every node near coord/.append style={font=\tiny,/pgf/number format/.cd,
            fixed,
            fixed zerofill,
            precision=1,
	    /tikz/.cd},
	nodes near coords, nodes near coords align={horizontal},
	legend style = {font=\scriptsize, legend columns=-1, /tikz/every odd column/.style={yshift=-1pt}},	
	legend entries={$\text{Output}_\text{HRTF-based}\quad\quad$, $\text{Output}_\text{Free-field-based}$},
	legend to name = legendResults
	]
	\addplot[black, fill=green] plot coordinates {(33.1,-10) (32.4,-5) (31.3,0) (32.2,5) (33.8,10)};
	\addplot[black, fill=red, postaction={pattern=north east lines}] plot coordinates {(40.1,-10) (38.7,-5) (37.9,0) (37.6,5) (36.8,10)};
      \end{axis}
    \end{tikzpicture}
    \label{fig:results_DOAError_Average_WER}
  }
  \hfill
  \subfigure[Average $\text{fwSegSNR}$.]{
    \begin{tikzpicture}
      \begin{axis}[width=0.24\textwidth, height=5cm, 
	label style = {font=\scriptsize},	
	tick label style = {font=\scriptsize},   
	ylabel style={yshift=-2.5mm}, 
	xbar=0pt,
	xmin=0, xmax = 18, ymin=-13.5, ymax=13.5,
	ytick=data,
	xtick={0, 5, 10, 15},
	xlabel = {$\text{fwSegSNR}\, [\text{dB}] \, \rightarrow$},
	bar width = 6pt, 
	every node near coord/.append style={font=\tiny,/pgf/number format/.cd,
            fixed,
            fixed zerofill,
            precision=1,
	    /tikz/.cd},
	nodes near coords, nodes near coords align={horizontal},
	]
	\addplot[black, fill=green] plot coordinates {(9.8,-10) (9.7,-5) (10.0,0) (9.6,5) (9.4,10)};
	\addplot[black, fill=red, postaction={pattern=north east lines}] plot coordinates {(8.8,-10) (9.0,-5) (9.1,0) (9.1,5) (8.9,10)};	
      \end{axis}
    \end{tikzpicture}
    \label{fig:results_DOAError_Average_FWSEGSNR}
  }
  \vspace{-3mm}
  \caption{Illustration of average \acp{WER} in $\%$ and average \ac{fwSegSNR} levels in $\text{dB}$, obtained at the output of the \ac{HRTF}- and free-field-based beamformer for $\phi_\mrm{d}=90^\circ$, $\phi_\mrm{int} \in \{10^\circ:(20^\circ):70^\circ, 110^\circ:(20^\circ):170^\circ\}$, and localization errors of $\pm 5^\circ$ and $\pm 10^\circ$. Average input measures: $\mathrm{WER}_\mrm{in}=47.1\%$ and $\mathrm{fwSegSNR}_\mrm{in}=5.5$dB.}
  \label{fig:results_DOAError_Average}
  \vspace{-5mm}
\end{figure}

In Figs.~\ref{fig:results_DOAError_Average_WER} and \ref{fig:results_DOAError_Average_FWSEGSNR}, the average \acp{WER} in $\%$ and \ac{fwSegSNR} levels in dB, obtained at the output of the \ac{HRTF}-based and free-field-based \ac{RLSFI} beamformer, respectively, are illustrated. The presented results were averaged over eight different scenarios, where the target source was always located at $\phi_\mrm{d}=90^\circ$ and the interferer was located at positions between $10^\circ$ and $70^\circ$, and $110^\circ$ and $170^\circ$, in steps of $20^\circ$. It can be seen that the average performance of the \ac{HRTF}-based beamformer decreases when there is a localization error. Furthermore, one can observe that the \ac{HRTF}-based beamformer in general yields a better performance than the free-field-based beamformer, as was already shown in \cite{Barfuss_WASPAA:2015}.
%
%

\subsection*{Impact of localization errors with respect to robot-source distance}
In a second experiment, we evaluated the impact of localization errors with respect to the robot-source distance $d_\mrm{RS}$. Since in our experiment the robot head and the source are not at the same height, distance errors result in a mismatch between the elevation angle the beamformer is steered to and the elevation angle of the target source with respect to the robot head array.
Here, we evaluated the beamformer performance for robot-source distances $d_\mrm{RS} \in \{1.1\text{m}, 2\text{m}\}$. The \ac{HRTF}-based beamformer was designed using \acp{HRTF} measured for a robot-source distance of $1.1$m. Thus, the elevation mismatch for $d_\mrm{RS} = 1.1$m, is $0^\circ$, whereas for $d_\mrm{RS} = 2\text{m}$, there is a mismatch of $13.5^\circ$, i.e., the beamformer is steered too high in elevation. To allow for a fair comparison, the same elevation angle was used for the free-field-based beamformer.

In Figs.~\ref{fig:results_DistanceError_Average_WER_a} and \ref{fig:results_DistanceError_Average_WER_b}, the average output \acp{WER} in $\%$ and \ac{fwSegSNR} levels in dB of the \ac{HRTF}- and free-field-based \ac{RLSFI} beamformers are illustrated. The results were obtained for the same desired and interfering source positions as for Fig.~\ref{fig:results_DOAError_Average}. 
The results show that a mismatch with respect to the robot-source distance leads to a significant decrease in \ac{WER} and to a slight decrease of the \ac{fwSegSNR}. 
Apart from that, it is interesting to see that the \ac{HRTF}-based beamformer still yields better results than the free-field-based beamformer.
\begin{figure}[t]
  {\hspace{15mm}\ref{legendResults}\vspace{-1.5mm}}  
  \subfigure[Average $\text{WER}$.]{
    \begin{tikzpicture}
      \begin{axis}[width=0.24\textwidth, height=3cm, 
	label style = {font=\scriptsize},	
	tick label style = {font=\scriptsize},   
	ylabel style={yshift=-1.5mm}, 
	xbar=0pt,
	xmin=0, xmax = 65, ymin=0.55, ymax=2.2,
	ytick={1, 1.75},
	yticklabels={$1.1$, $2$},
	xtick={0, 20, 40, 60},
	ylabel = {$d_\mrm{RS}\,[\text{m}] \, \rightarrow$},
	xlabel = {$\text{WER}\, [\%] \, \rightarrow$},
	bar width = 6pt, 
	every node near coord/.append style={font=\tiny,/pgf/number format/.cd,
            fixed,
            fixed zerofill,
            precision=1,
	    /tikz/.cd},
	nodes near coords, nodes near coords align={horizontal},
	legend style = {font=\scriptsize, legend columns=-1, /tikz/every odd column/.style={yshift=-1pt}},	
	legend entries={$\text{Output}_\text{HRTF-based}\quad\quad$, $\text{Output}_\text{Free-field-based}$},
	legend to name = legendResults
	]
	\addplot[black, fill=green] plot coordinates {(31.4,1) (36.6,1.75)};
	\addplot[black, fill=red, postaction={pattern=north east lines}] plot coordinates {(37.9,1) (40.5,1.75)};
      \end{axis}
    \end{tikzpicture}
    \label{fig:results_DistanceError_Average_WER_a}
  }
  \hfill
  \subfigure[Average $\text{fwSegSNR}$.]{
    \begin{tikzpicture}
      \begin{axis}[width=0.24\textwidth, height=3cm, 
	label style = {font=\scriptsize},	
	tick label style = {font=\scriptsize},   
	ylabel style={yshift=-2.5mm}, 
	xbar=0pt,
	xmin=0, xmax = 18, ymin=0.55, ymax=2.2,
	ytick={1, 1.75},
	yticklabels={$1.1$, $2$},
	xtick={0, 5, 10, 15},
	xlabel = {$\text{fwSegSNR}\, [\text{dB}] \, \rightarrow$},
	bar width = 6pt, 
	every node near coord/.append style={font=\tiny,/pgf/number format/.cd,
            fixed,
            fixed zerofill,
            precision=1,
	    /tikz/.cd},
	nodes near coords, nodes near coords align={horizontal},
	]
	\addplot[black, fill=green] plot coordinates {(10.0,1) (9.8,1.75)};		
	\addplot[black, fill=red, postaction={pattern=north east lines}] plot coordinates {(9.1,1) (8.8,1.75)};	
      \end{axis}
    \end{tikzpicture}
    \label{fig:results_DistanceError_Average_WER_b}
  }
  \vspace{-3mm}
  \caption{Illustration of average \acp{WER} in $\%$ and \ac{fwSegSNR} levels in dB, obtained at the output of the \ac{HRTF}- and free-field-based beamformer for $\phi_\mrm{d}=90^\circ$, $\phi_\mrm{int} \in \{10^\circ:(20^\circ):70^\circ, 110^\circ:(20^\circ):170^\circ\}$, and robot-source distances of $1.1$m and $2$m. Average input measures: $\mathrm{WER}_\mrm{in}=47.3\%$ and $\mathrm{fwSegSNR}_\mrm{in}=5.5$dB.} 
  \label{fig:results_DistanceError_Average_WER}
  \vspace{-5mm}
\end{figure}

\section*{Conclusion}
\label{sec:summary_conclusion}
In this work, we investigated the impact of localization errors on the performance of a recently proposed \ac{HRTF}-based \ac{RLSFI} beamformer. Localization errors with respect to the \ac{DOA} of the target signal as well as the robot-source distance were evaluated. The results confirmed that both, erroneous \ac{DOA} and robot-source distance estimates lead to a significant decrease in beamforming performance. Thus, it is of vital importance to use a set of \acp{HRTF} for the design of the \ac{HRTF}-based \ac{RLSFI} beamformer, which matches the position of the target source.
Future work includes analysis of the effect of localization errors on the behaviour of the \ac{HRTF}-based beamformer for sources in the near-field, and an extension 
of the \ac{RLSFI} beamformer design to two-dimensional beam steering.


\end{document}